\documentclass[preprint,superscriptaddress,prl]{revtex4-1}
\usepackage{graphicx}
\usepackage{amsmath,amssymb}
\usepackage{mathrsfs}
\usepackage[normalem]{ulem}
\usepackage{color}
\usepackage{soul}
\usepackage{hyperref}
\usepackage[T1]{fontenc}
\usepackage[utf8]{inputenc}
\usepackage{bm}
\usepackage[left]{lineno}

\newcommand{\be}{\begin{equation}}
\newcommand{\ee}{\end{equation}}
\newcommand{\bea}{\begin{eqnarray}}
\newcommand{\eea}{\end{eqnarray}}

\newcommand{\SI}{Supplementary Information}
\newcommand{\vbg}{V_{{\rm BG}}}
\newcommand{\vsd}{V_{{\rm SD}}}
\newcommand{\Isd}{I_{{\rm SD}}}
\newcommand{\mo}{\rm MoS_{{2}}}
\newcommand{\ignore}[1]{}
\newcommand{\UFMG}{Departamento de Fisica, Universidade Federal de Minas Gerais, Belo Horizonte, MG 31270-901, Brazil}
\newcommand{\NIMS}{Advanced Materials Laboratory, National Institute for Materials Science, 1-1 Namiki, Tsukuba 305-0044, Japan}

\renewcommand{\phi}{\varphi}
\renewcommand{\epsilon}{\varepsilon}

\usepackage{fancyhdr}
\begin{document}

\title{Gate-tunable non-volatile photomemory effect in $\mo$ transistors}

\author{Andreij C. Gadelha}
\affiliation{\UFMG}

\author{Alisson R. Cadore}
\affiliation{\UFMG}

\author{Kenji Watanabe}
\affiliation{\NIMS}

\author{Takashi Taniguchi}
\affiliation{\NIMS}

\author{Ana M. de Paula}
\affiliation{\UFMG}

\author{Leandro M. Malard}
\affiliation{\UFMG}

\author{Rodrigo G. Lacerda}
\affiliation{\UFMG}

\author{Leonardo C. Campos}
\affiliation{\UFMG}

\begin{abstract}
\date{\today}

\textbf{Non-volatile memory devices have been limited to flash architectures that are complex devices. Here, we present a unique photomemory effect in $\bm{{\mo}}$ transistors. The photomemory is based on a photodoping effect - a controlled way of manipulating the density of free charges in monolayer $\bm{{\mo}}$ using a combination of laser exposure and gate voltage application. The photodoping promotes changes on the conductance of $\bm{{\mo}}$ leading to photomemory states with high memory on/off ratio. Such memory states are non-volatile with an expectation of retaining up to 50\% of the information for tens of years. Furthermore, we show that the photodoping is gate-tunable, enabling control of the recorded memory states. Finally, we propose a model to explain the photodoping, and we provide experimental evidence supporting such a phenomenon. In summary, our work includes the $\bm{{\mo}}$ phototransistors in the non-volatile memory devices and expands the possibilities of memory application beyond conventional memory architectures.}
\end{abstract}

\maketitle
\thispagestyle{fancy}

\subsection*{Introduction}
\lhead{}
Due to the ultra-thin nature and tunable electrostatic properties of two-dimensional (2D) materials, they have strategical importance for digital electronics and memory applications. The required features (figures of merit) for actual memory devices include the miniaturisation capability, the power consumption (high memory on/off ratio), the operation speed, the memory retention time and the cost. Innately, the 2D materials based devices show advantages for miniaturization, however to date there are no reported simple memory devices that cover most of the required features, notably the memory retention time~\citep{mgmemo,tuncha,nonme,lpnc,monoopt,therass,multires,grflash,grferr,gramen,grbis,multigraphene,pol,memopn,float1,float2,float3}.
Monolayer $\mo$ is a direct band gap 2D semiconductor material \cite{slmt,2dreview} that shows high photocurrent response \cite{ultra,ultraviolet,cvdfot,slmp,hdm,pnreview}, high photoluminescence emission \cite{lummos2,eplm} and interesting valleytronic properties \citep{vpmm,cvpm,vscd,tvhe}. Thus, $\mo$ based flash memory devices with high memory on/off ratio and long memory retention time have emerged. However, the implementation of flash devices is challenging because they require engineering with many elements into complex architectures. More recently, floating-gate tunneling devices using simpler architechtures than flash devices have been proposed but they lack ultrahigh time-stability \cite{float1,float2,float3}. Therefore, the development of alternative, high-performance, simpler memory architectures is strategical.  
Toward this direction, some reports have investigated a thermally assisted memory effect \cite{therass} and an optical memory effect \cite{monoopt} in a $\mo$ field effect transistor (FET) that is also a simpler architecture than flash devices. Nonetheless, in the first case, the memory effect has a drawback of not operating at room temperature and both cases \cite{therass,monoopt} showed low memory on/off ratio and short memory retention time. 

Here we show that we can obtain a non-volatile photomemory effect with high on/off ratio in a $\mo$ FET architecture.  Such photomemory effect is based on a photodoping process that changes the $\mo$ conductance in a way that promotes two distinguishable binary photomemory states with on/off ratio up to 10$^{6}$. The photogenerated memory states are persistent and predicted to retain up to 50\% of its information for decades, that leads to a non-volatile photomemory. Moreover, it is important to mention that the presented photomemory is gate-tunable. The gate voltage is used to both adjust the memory on/off ratio (with the laser off) and to manipulate the recorded photomemory states during the laser exposure. Finally, we explore and discuss a possible physical mechanism of the photodoping that is also supported by our experimental evidence. In summary, we propose a photomemory effect in $\mo$ FETs that expands the possibilities of memory application beyond conventional memory architectures.

\subsection*{Results and Discussion}

\begin{figure*}[!hbtp]
\centering
\centerline{\includegraphics[width=17.4cm]{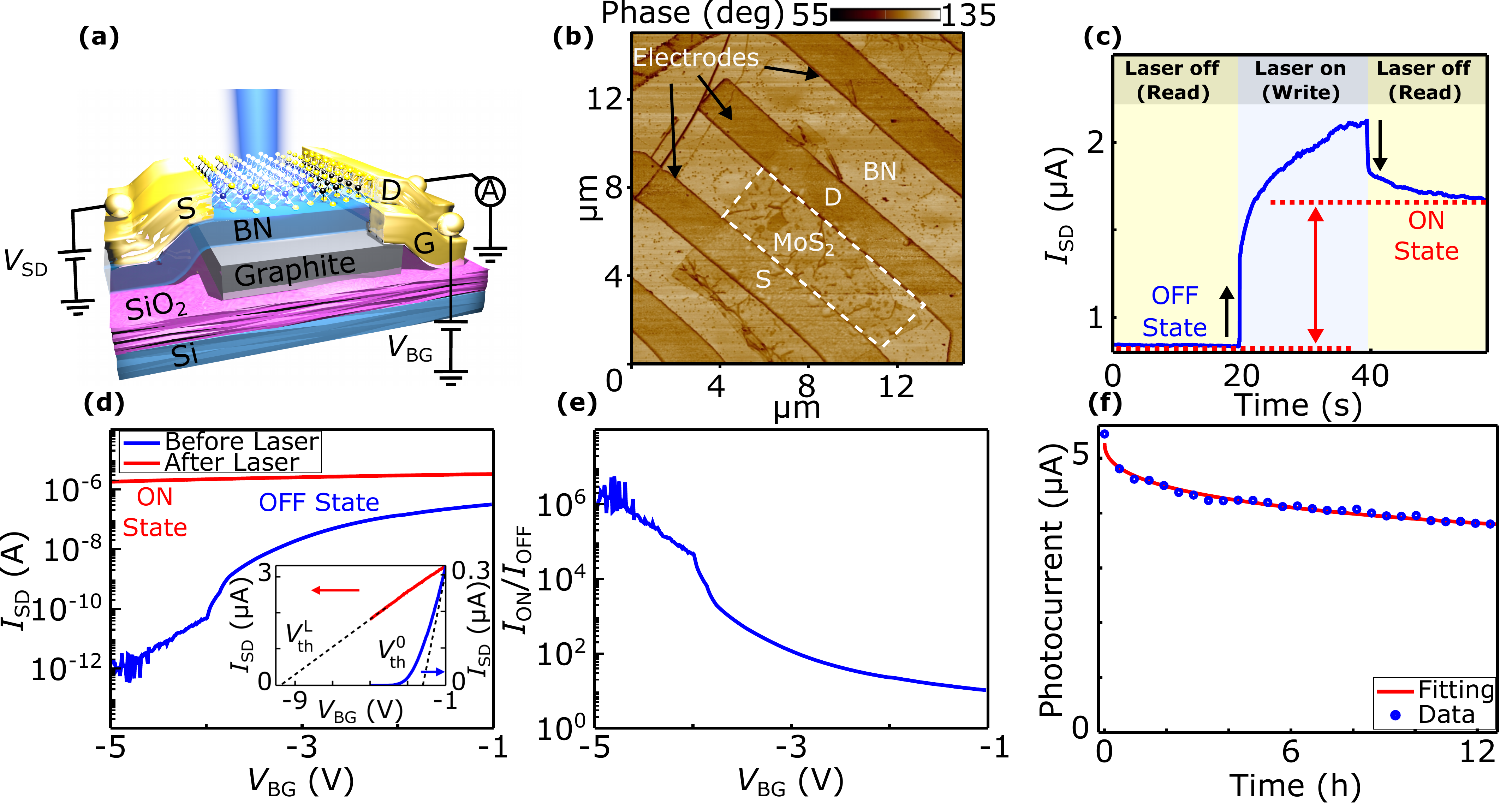}}
\caption{\label{fig:1}\textbf{Photodoping and non-volatile photomemory}. \textbf{(a)}, sketch of the $\mo$ FET. \textbf{(b)}, AFM phase image of the device. \textbf{(c)}, time resolved photocurrent, laser exposure at 488~nm with fluence of 60~$\mu$W$/\mu$m$^{2}$. The parameters are $\vbg =\rm{-5\,V}$ and $\vsd =\rm{0.1\,V}$. \textbf{(d)}, The $\Isd$ vs $\vbg$ measurements on a log scale before (blue) and after (red) the laser exposure, $V_{\mathrm{SD}}=\mathrm{0.1\,V}$. In the inset, the same measurements but on a linear scale. The red curve is measured after the 488~nm laser exposure with fluence of 700~$\mu$W/$\mu$m$^{2}$ and $\vbg=\mathrm{-5\,V}$ until photocurrent saturation. \textbf{(e)}, $I_{\mathrm{ON}}/I_{\mathrm{OFF}}$ ratio as a function of the gate voltage. \textbf{(f)}, photocurrent decay after the photodoping induced by the 488~nm laser with fluence of 700~$\mu$W/$\mu$m$^{2}$ until photocurrent saturation. The parameters $V_{\rm{BG}}=\rm{0\,V}$ and $V_{\rm{SD}}=\rm{0.1\,V}$ are used for this measurement.}
\end{figure*}

The photomemory effect investigated in this work is due to the modulation of the conductance of a monolayer $\mo$ field effect transistor via a simultaneous application of light and electrostatic gate potential. Along this paper, we show evidence that the main mechanisms for the photomemory relies on the manipulation of a charging effect at the gate-insulator interface of the FET (the interface between the insulator and the material of the gate terminal). Although other mechanisms can have some influence on the photomemory, we show that our model explains well our results. Our FET is a Van der Waals heterostructure consisting of a monolayer $\mo$ supported by a high-quality hexagonal Boron Nitride crystal (BN), see Fig. \ref{fig:1}(a). In this case, we use a graphite crystal to provide a flat back gate electrode.  In  Fig. \ref{fig:1}(b) we present an atomic force spectroscopy (AFM) phase image of one of the devices measured in this work. While, in the Supplementary Information we depict the characterization for the second device.

We start by presenting the process of photocurrent generation in the $\mo$ FET. Fig. \ref{fig:1}(c) shows a typical time-resolved photocurrent measurement of our device. Initially we measure the standard current ($I\rm{_{SD}}$) in dark conditions, then we illuminate the device using the laser ($\lambda=\mathrm{488\,nm}$) for $\mathrm{20\,s}$ and $\vbg=\mathrm{-5\,V}$. We use the same laser with $\lambda=\mathrm{488\,nm}$ for all the optoelectronic measurements that we present in this text. A careful analysis of the current as a function of time reveals that there are two optical processes generating the photocurrent in the $\mo$ channel. First, we observe a rapid increase of $\Isd$ due to excitation of electron-hole pairs (see vertical black arrow), then a second and slow process that dominates the photocurrent. We observe the same trend when the laser is turned off. There is a rapid collapse of $\Isd$, due to the recombination of electron-hole pairs, then a prolonged decay process that leads to a persistent photocurrent (PPC). The photodoping effect causes the PPC. We will discuss this process later. For now, we will define the photomemory states ``ON'' and ``OFF''. In Fig. \ref{fig:1}(c) we ascribe the PPC as the ``ON'' state, while the current before the laser exposure is the ``OFF'' state, which are binary photomemory states.

The photomemory effect is better observed in Fig. \ref{fig:1}(d), where we show the transfer curves, $\Isd$ vs back gate voltage ($\vbg$) curves, on a log scale. In the inset, we plot the the same curves, but on a linear scale. In blue, we plot a transfer curve before the laser exposure. By extrapolating the $\Isd$ curve we estimate the threshold voltage ($V\rm_{th}$) as $V\rm_{th}^{0}=\rm{-2.2\,V}$, see the inset in Fig. \ref{fig:1}(d). After this, the $\mo$ device is exposed to the laser beam with $\vbg=\mathrm{-5\,V}$ until the photocurrent saturates. The reason for waiting the photocurrent saturation is to reach the best response of our device. Then, we turn the laser off and repeat the transfer curve measurement. The data from the transfer curve after the laser exposure (red curve in Fig. \ref{fig:1}(d)) displays a significant increase of $\Isd$ at all applied gate conditions. It is important to note that there is a shift of $V\rm_{th}$ towards $\vbg$ out of the range of the experiment. This shift is a signature of a photodoping effect. It means that the density of free charges of $\mo$ has changed after the laser exposure. We estimate by extrapolating the data that the initial $V\rm_{th}^{\rm{0}}=\rm{-2.2\,V}$ shifts to $V\rm_{th}^{L}=\rm{-9.8\,V}$, see the inset in Fig. \ref{fig:1}(d). Also, the expected change in the density of charge of the $\mo$ due to photodoping is $\Delta n_{\textrm{ph}}=\mathrm{6\times 10^{12}\,cm^{-2}}$. Which is evaluated using the equation:

\begin{equation} \label{eq:n}
 \Delta n_{\textrm{ph}}=\frac{\epsilon_{0}\epsilon_{ox}}{e\,d}(V\rm_{th}^{L}-V\rm_{th}^{0})
\end{equation} 
where $\epsilon\rm_{ox}$ and $d$ are the dielectric constant of the insulator and its thickness, respectively. Note that such extra doping is obtained simply by the combination of the laser exposure and the applied gate bias. 

We now describe the methods used here to define the photomemory states, that can also be used to perform the ``read'' operations. We consider as an ``OFF'' state the measured $\Isd$ before the laser exposure for a given $\vbg$ (no information is recorded in the photomemory), see the blue curve in Fig. \ref{fig:1}(d). Similarly, the measured $\Isd$ after the laser exposure for the same $\vbg$ is considered as an ``ON'' state, see the red curve Fig. \ref{fig:1}(d). Another method to determine, or to ``read'', the photomemory states is by measuring $\Delta n_{\textrm{ph}}$ before and after the laser exposure. We reinforce that we perform the ``read'' operations with the laser off. We will discuss the ``record'' operations later, which are the procedures that ``write'' and ``erase'' the memory states. Because $\Isd$ is a function of $\vbg$, by measuring $\Isd$ instead of $\Delta n_{\textrm{ph}}$ we have the advantage to use the gate voltage (with the laser off) to optimise the gain of the photomemory \cite{mgmemo,pdh}. We elucidate this fact in Fig. \ref{fig:1}(e), where we plot the $I_{\rm{ON}}/I_{\rm{OFF}}$ ratio (memory on/off ratio) as function of $\vbg$.  The $I_{\rm{ON}}/I_{\rm{OFF}}$ ratio changes from 10, for positive gate voltages, to values up to 10$^{6}$, for negative gate voltages. Observe that the high modulation of the memory on/off ratio with gate voltage is an attribute of the photomemory effect. It must be noted that to reach the photocurrent saturation we do exposures of 30~min, for example in Fig. \ref{fig:1}(d). However, we can also obtain a high memory on/off ratio of 10${^4}$ with a short exposure time (20~s), see Fig. S14.

Another crucial figure of merit of a memory device is the memory retention time. To assess that, we measure the $\mo$ photocurrent decay over time, applying $\vsd=\mathrm{0.1\,V}$ and $\vbg=\mathrm{0\,V}$, see the blue dots in Fig. \ref{fig:1}(f). We measure the decay after the photocurrent saturation by the laser exposure. After $\mathrm{15\,h}$ the photocurrent barely decreases, suggesting that the photomemory state is permanent. So, the photomemory is a non-volatile memory. To estimate the memory loss over ten years, we employ an exponential decay fit, the red line in Fig. \ref{fig:1}(f). From the fitting, we predict that the reminiscent memory current for the photomemory device is approximately 50\% of the initial photocurrent. Thus, the devices can retain 50\% of the memory for ten years. These values are much better than the ones for the $\mo$ flash memory architectures, where the retention percentage is in the range of 15-30\% \cite{mgmemo,nonme,tuncha,lpnc}. 

We can now describe in more details the photomemory device, which is composed mainly of two elements in the FET architecture. One element is the gate-insulator interface, where possibly the charges are trapped inducing the photodoping. The other element is the semiconductor channel, from which we ``read'' the photomemory states. In this way, we can design better photomemory devices by choosing other gate-insulator interfaces that can provide higher values of photodoping and retention time. Furthermore, the choices of semiconductors with better mobility and subthreshold swing would enable to achieve higher on/off ratio values. 

\begin{figure*}[!hbtp]
\centering
\includegraphics[width=11.36cm]{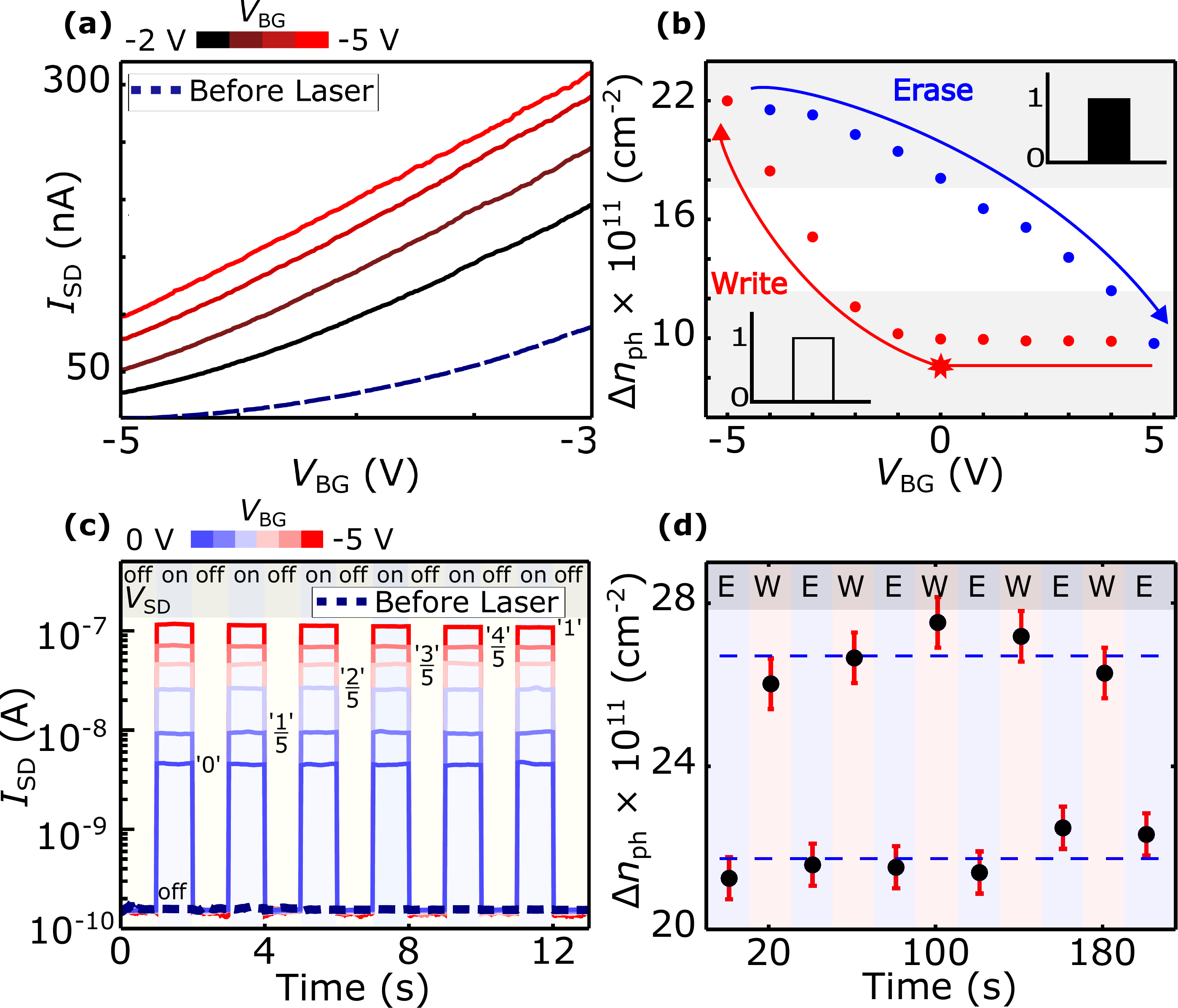}
\caption{\label{fig:2}\textbf{Gate-tunable photomemory}. \textbf{(a)}, $\Isd$ vs $\vbg$ curves before laser exposure, blue curve, and after laser exposures with $\vbg$ values defined in the color bar. \textbf{(b)}, $\Delta n_{\textrm{ph}}$ vs $V_{\rm{BG}}$ curve. First the point $V_{\rm{BG}}=\rm{0\,V}$ is measured and then the arrows indicates the followed applied gate voltages during the laser exposures. For figures \textbf{(a)}-\textbf{(b)} there is a 488~nm laser exposure on each point for 20~s (at laser fluence of 60~$\mu$W/$\mu$m$^{2}$). \textbf{(c)}, multilevel photomemory, gate values from 0~V to $-$5~V are used for the ``writings'' and 20~s of laser exposure at fluence of 700~$\mu$W/$\mu$m$^{2}$. For the ``readings'' a gate value of $-$4~V and bias pulses of 0.1~V are applied. \textbf{(d)}, ``write''-``erase'' operations, for $\vbg=\mathrm{-5\,V}$ and $\vbg=\mathrm{5\,V}$, respectively, and 20~s of laser exposure (fluence of 700~$\mu\mathrm{W}/\mu\mathrm{m^{2}}$).}
\end{figure*}

The photomemory achieved on a FET architecture has the advantage that we can improve some features by choosing the adequate gate voltages. One example is the high memory on/off ratio already discussed. Another important feature is that we can select distinguishable photomemory states for the same laser exposure due to the photodoping dependence on the gate voltage. In this way, the gate voltage is used both to ``read'' and to ``record'' the photomemory states. Here, we define of ``record'' operation the procedure of doing in our devices laser exposures concomitantly with the gate voltage application. In Fig. \ref{fig:2} we show results that highlight the ``recording'' of the photomemory states. Fig. \ref{fig:2}(a) reveals the changes in the density of free charges due to the photodoping effect by exhibiting multiple transfer curves at different photomemory states. The blue curve represents the ``OFF'' state before any laser exposure. After the ``OFF'' state is measured, by evaluating a transfer curve of the device, we ``record'' a photomemory state by applying $\vbg=\mathrm{-2\,V}$ and by exposing the photomemory device to the laser for 20~s. After this ``record'' operation, we measure a new transfer curve with the laser off (black curve in Fig. \ref{fig:2}(a)) and from the data of Fig. \ref{fig:2}(a) we observe that the $\mo$ sheet acquires a new density of charge after the laser exposure. We evaluate the new density of free charges from the equation~1 as a function of the new $V\rm_{th}$. To visualise how the transfer curve changes at every ``record'' operation, the process described above is repeated applying gate voltages during the ``record'' operations up to $\vbg=\mathrm{-5\,V}$ in steps of $\mathrm{-1\,V}$, as shown in Fig. \ref{fig:2}(a). It is interesting to note that for each ``record'' operation with different $\vbg$ there is a distinct transfer curve and thus a particular photomemory state. Then, we can choose several ``ON'' states with distinct electrical conductivity.     

Fig. \ref{fig:2}(b) exposes the change of the density of free charges acquired for the $\mo$ after every photomemory ``record'' operation as compared to the intrinsic density of charge of the photomemory (equation~1). Here we use the same ``record'' operation described in Fig. \ref{fig:2}(a), but we achieve the initially recorded state by applying a $\vbg=\mathrm{0\,V}$ during the 20~s of laser exposure. We name this recorded state as ``0'' state. We ``record'' the other photomemory states in the arrows indicated sequence by changing the gate voltages in a range of $\mathrm{-5\,V \leq\vbg \leq 5\,V}$.
The negative gate voltages are used during the ``record'' operations to monotonically increase the density of charge to set a ``1'' state and the positive gate voltages are used during the ``record'' operations to reduce the density of charge and to restore the initial ``0'' state.  We name the process of charge injection in the $\mo$ as a ``write'' operation (red arrow). We perform the ``write'' operation by exposing the device to the laser with an applied negative gate voltage. We denominate the process of removing the charges as an ``erase'' operation (blue arrow). The gate-``erase'' operation is performed right after writing the ``1'' state, but doing ``record'' operations with gate voltages larger than $\vbg=\mathrm{-5\,V}$. For example, in fig. \ref{fig:2}(b) the ``erase'' operations are executed with several laser exposures applying $\vbg=\mathrm{-4\,V}$,$\mathrm{-3\,V}$ ... $\mathrm{5\,V}$. Also, note that the ``erase'' operation cannot cancel the photodoping completely, so we still have a reminiscent photodoping of $\Delta n_{\textrm{ph}}=\mathrm{10\times 10^{11}}$ after the ``erase'' operation, see Fig. \ref{fig:2}(b). 

Fig. \ref{fig:2}(b) also shows that the laser ``record'' operations with different gate voltages generate distinct $\Delta n_{\textrm{ph}}$ values, which correspond to distinct photomemory states. This dependence of the photomemory states on the $\vbg$ used during the ``record'' operations shows that we can use the photomemory for multilevel ``ON'' memory states operation. However, we must point out that it is not the aim of this work to explore multilevel memory operation. We only elucidate that the gate-tunability property of photomemory can allow this type of operation. Fig. \ref{fig:2}(c) demonstrate how this is possible employing current readings. In Fig. \ref{fig:2}(c), the dashed black line represents the ``OFF'' state, which is measured by applying $\vsd=\mathrm{0.1\,V}$ and $\vbg=\mathrm{-4\,V}$ before any laser exposure. We ``record'' ``ON'' states applying laser exposures using different $\vbg$ at each ``record'' operation. More precisely, we use $\vbg$ from 0~V to $-$5~V, with increments of $-$1~V, and do laser exposures for 20~s, ``recording'' multilevel states which we denote by ``0'', ``$\mathrm{\frac{1}{5}}$'', ``$\mathrm{\frac{2}{5}}$'' ... ``1''. After each ``record'' operation, we ``read'' the photomemory state by measuring the current through the device at the same electrostatic condition used when we ``read'' the ``OFF'' state. The difference here is that we use pulses of $\vsd=\mathrm{0.1\,V}$ for 2~s spaced by 2~s to show that the information is stored in the photomemory even when no $\vsd$ is applied. Note that the ``OFF'' state is shown only for reference and the ratio between ON/OFF states previously discussed does not apply for multilevel operations. However, the gain between such multilevel states can be maximised tuning the gate-potential, but in our presented data it is of the order of ten.  

Although multilevel memory states are interesting, here they are explored only to demonstrate the usefulness of the gate-tunability property of the photomemory. However, for practical memory operations, it is straightforward to explore the reliability in the ``write''-``erase'' operations between the binary memory states. In this case, we generate the binary ``1'' and ``0'' states by applying $\vbg=\mathrm{-5\,V}$ and $\vbg=\mathrm{5\,V}$, respectively, during the laser exposures of 20~s. We show the reproducibility and reliability of the ``write''-``erase'' operations of the binary memory states in Fig. \ref{fig:2}(d), that presents a sequence of successful ``write''-``erase'' cycles. These results demonstrate the device robustness. In Fig. \ref{fig:2}(d), it is also represented the error bars in each ``record'' operation. The error bars show that the ``write''-``erase'' operations generate distinguishable photomemory states.

It is worth mentioning that the variation of the photodoping between the ``1'' and the ``0'' states in Fig. \ref{fig:2}(b) is $\Delta n_{\rm{ph}}\sim\mathrm{10^{12}\,cm^{-2}}$, which is evaluated by $\Delta n_{\rm{ph}}=\frac{\epsilon_{0}\epsilon_{ox}}{e\,d}(V\rm_{th}^{\text{``1''}}-V\rm_{th}^{\text{``0''}})$, where $V\rm_{th}^{\text{``1''}}$ and $V\rm_{th}^{\text{``0''}}$ are the threshold voltage of the device in the ``1'' and in the ``0'' states, respectively. Recall that to obtain this modulation of the photodoping we do laser exposures of 20~s together with gate voltage applications. The generated photodoping with a 20~s laser exposure is an order of magnitude lower than the maximum photodoping ($\Delta n_{\rm{ph}}\sim\mathrm{10^{13}\,cm^{-2}}$) obtained in this work, see Fig. \ref{fig:1}(d), where the photodoping is maximized by waiting for the saturation of the photocurrent after 30~min laser exposure. Such high photodoping modulation give an ultra-high memory on/off ratio of $10^{6}$ in Fig. \ref{fig:1}(e). However, we can still obtain a high memory on/off ratio of $10^{4}$ by using laser exposures of 20~s, see Fig. S14.

Finally, we discuss the process of photodoping that possibly generates the PPC and the photomemory effect in our $\mo$ FETs. It is important to mention that the PPC is not a consensus topic. The most discussed explanations for the PPC in $\mo$ is either due to the photo-induced charge transfer from adsorbed gases to the $\mo$ channel \cite{pmf} or due to the Coulomb interaction with defects at the insulator surface \cite{ultra,ultraviolet,mpat,cvdfot,extrinsic,pdh,fepn,etpp}. We believe that the interactions with adsorbed gases are not a valid explanation in our devices as there is no hysteresis in the $\sigma$ vs $V_{\rm{BG}}$ curves when we sweep the voltage in opposite directions \cite{hysteresis}(see Fig. S7). We believe that the interactions with defects at the insulator surface are not the dominant mechanism, as the devices have a low density of defects when compared with the photodoping observed in our work (10$^{13}$~cm$^{-2}$). Furthermore, the fact that we measure the photodoping in a clean and flat BN substrate \cite{BN} reinforces this statement. Consider that we use a $\mathrm{\sim30\,nm}$ thick BN, which prevents tunneling as a charge-trapping mechanim like occurs in the reference \cite{float1,float2,float3}. It should be also mentioned that to prepare our Van der Waals heteroestructures we use the same wet-transfer method of the reference \cite{BN}, which leaves some bubbles and wrinkles between the BN and the graphite flakes, see Fig. S1 and Fig. S2. However, in spite of these issues, the BN is clean and atomically flat in the majority of the surface of the devices. Recall that we do not study the influence of these imperfections between the layers in the photodoping effect, but we do not discard that they can play a role.

\begin{figure*}[btp]
\centering
\includegraphics[width=11.2cm]{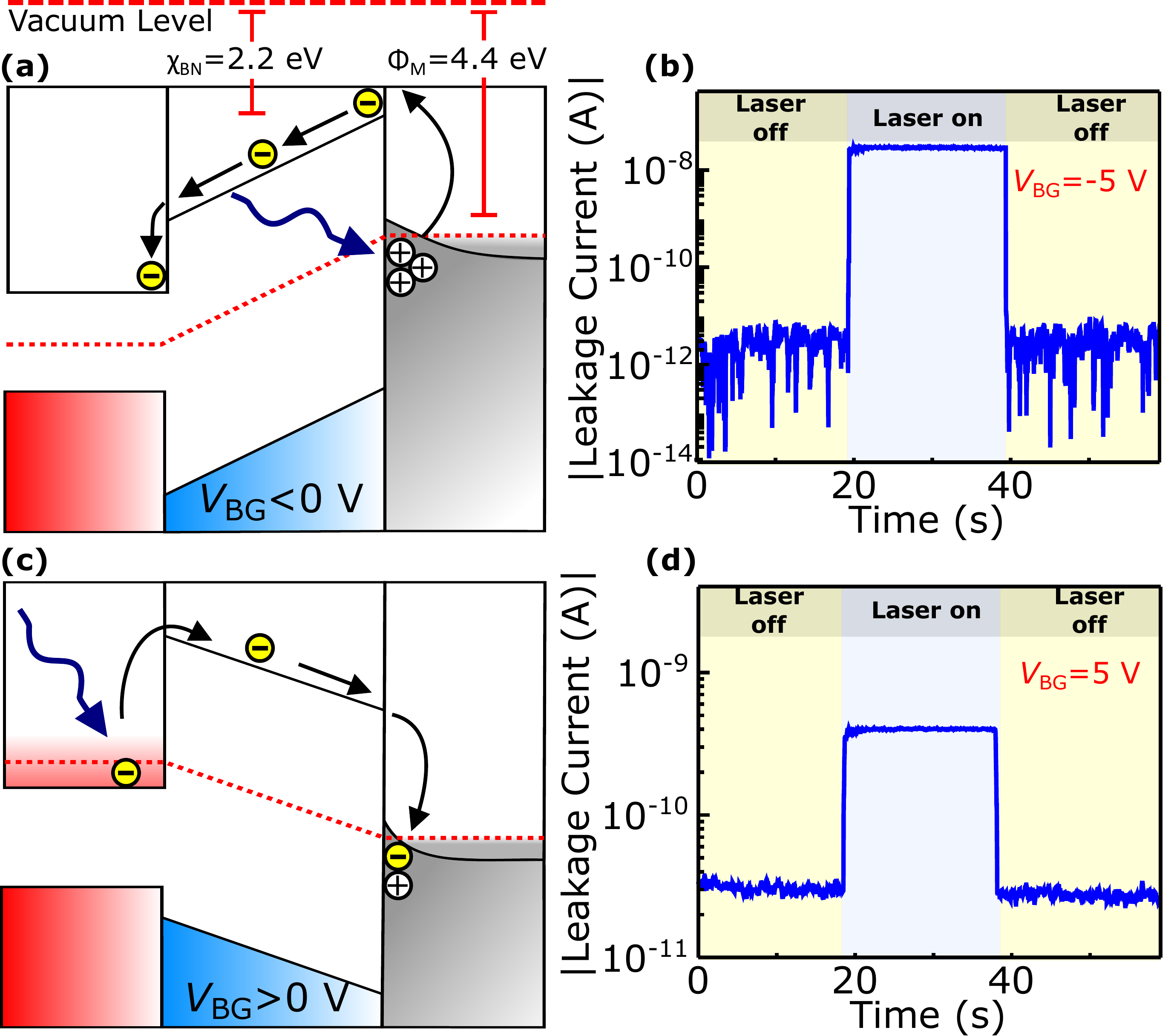}
\caption{\label{fig:4}\textbf{Physical model for the photodoping}. \textbf{(a)}, energy band diagram for the $\mo$/BN/graphite junction as a function of position for $\vbg\mathrm{<0\,V}$. \textbf{(b)}, photogenerated leakage current, $\vbg\mathrm{=-5\,V}$, $\lambda=\mathrm{488\,nm}$ and fluence of 60~$\mathrm{\mu W/\mu m^{2}}$. \textbf{(c)}, energy band diagram for the $\mo$/BN/graphite junction as a function of position for $\vbg\mathrm{>0\,V}$. \textbf{(d)}, photogenerated leakage current, $\vbg\mathrm{=5\,V}$, $\lambda=\mathrm{488\,nm}$ and fluence of 60~$\mathrm{\mu W/\mu m^{2}}$.}
\end{figure*}

Thus, we shall attribute a different process to the photodoping in our $\mo$ FET, which we propose to be a photogeneration of trapped holes in the gate-insulator interface. We clarify this mechanism by drawing the energy band diagram of the $\mo$ FET. Fig. \ref{fig:4}(a) shows a band diagram of the $\mo$ FET with $\vbg<\mathrm{0\,V}$ applied to the graphite relative to the $\mo$. Here $\chi_{\mathrm{MoS}_{2}}$ is the $\mo$ electron affinity, $\chi\mathrm{_{BN}}$ is the BN electron affinity and $\Phi_{\rm{M}}$ is the graphite work function. We also show in the gate-insulator junction the bending of the graphite band, that generates a built-in electric field. For $\vbg<\mathrm{0\,V}$, photons with sufficient energy ($E_{\rm{L}}>\Phi_{\rm{M}}-\chi\mathrm{_{BN}}$) promote the electrons from the gate-insulator interface to the conduction band of BN. The applied negative gate voltage drives these photoexcited electrons through the conduction band of BN to the $\mo$ channel, but some holes generated during the photoabsorption process remain trapped at the gate-insulator interface by the electric field of the gate-insulator junction. The positively charged layer generates photodoping in the $\mo$ channel, see Fig. \ref{fig:4}(a). According to this energy diagram description, we predict that we should observe a photo-generated leakage current under laser exposure between the drain and gate electrodes. This fact is verified in our experiments, as depicted in Fig. \ref{fig:4}(b) that exhibits a 10$^{-8}$~A leakage current during the laser exposure.

For $\vbg>\mathrm{0\,V}$, the $\mo$ channel is n-doped, so when we turn the laser on, the electrons from $\mo$ are photoexcited to the conduction band of BN, see Fig. \ref{fig:4}(c). In this case, the gate-field drives these electrons through the conduction band of BN to the gate-insulator junction, recombining with some of the trapped holes, reducing the photodoping. We do not achieve the photodoping reduction process totally, because the built-in electric field of the gate-insulator junction prevents some of the electrons to recombine. Fig. \ref{fig:4}(d) shows that for $\vbg>\mathrm{0\,V}$ we can also observe a photo-generated leakage current during the laser exposure. Note that for $\vbg>\mathrm{0\,V}$ the photo-generated leakage current is lower than for $\vbg<\mathrm{0\,V}$. We can associate this fact to the density of states of $\mo$, which is smaller than the graphite flake. It is also important to mention that this asymmetry in the photo-generated leakage current imposes a faster ``write'' operation relative to the ``erase'' operation, see Fig. S10.

The proposed model in Fig. \ref{fig:4} explains the results of Fig. \ref{fig:2}(b), which shows that the applied negative gate bias increase the photodoping, whether positive gate bias reduce the photodoping. Moreover, the threshold energy for photodoping generation ($E_{\rm{th}}=\Phi_{\rm{M}}-\chi\mathrm{_{BN}}$) in Fig. \ref{fig:4} matches our experimental results. Indeed, a crystal of BN possess a band gap ($E_{\rm{g}}^{\rm BN}$) of 5.2-5.9~eV and electron affinity ($\chi_{\rm{BN}}$) of 2.0-2.3~eV \cite{mgmemo}. Whereas graphite has a work function ($\Phi_{\rm{G}}$) of 4.3-4.6~eV \cite{tcog,dies,cwfg}. Therefore, the difference between $\Phi_{\rm G}$ and $\chi_{\mathrm{BN}}$ is around 2.2~eV, so only photons with energy larger than $\mathrm{2.2\,eV}$ are predicted to promote photoexcitation, see Fig. \ref{fig:4}(a). Therefore, we have done measurements with a laser energy of 1.6~eV and measured an almost negligible photodoping of $\Delta n_{\textrm{ph}}\sim\mathrm{10^{10}\,cm^{-2}}$ (see Fig. S11). In contrast, for the laser energy of 2.5~eV we have observed a high photodoping of $\Delta n_{\textrm{ph}}\sim\mathrm{10^{12}\,cm^{-2}}$ (see Fig. S13). The small, but not null, photodoping with the 1.6~eV laser may be due to other minor effects that may also occur, as the excitation of defects from the $\mo$ channel \cite{STMd}. However, mostly the gate-insulator interface contains the physics of the photodoping, therefore studying other materials may enable photomemory improvements.

\subsection*{Conclusion}
In conclusion, we have demonstrated that it is possible to obtain a non-volatile photomemory effect with high on/off ratio in a FET architecture. We showed that high values of doping are achieved via laser exposure that generates the binary photomemory states with high on/off ratio. We have shown that the photomemory described presents long memory retention time and thus the photomemory states are non-volatile. We have also verified that the photomemory states can be controlled and adjusted by the applied gate voltage, that could also be used to improve the memory on/off ratio. Finally, we have proposed a phenomenological model that agrees well with the experimental observations and clarifies a possible nature of the photodoping effect in $\mo$ FETs. Our results widen the possibilities of memory applications using 2D materials.

\section*{Methods}

\textbf{Device Fabrication}. The devices are obtained by transfer \cite{BN} of BN crystals ($\sim\mathrm{30\,nm}$ thick) to graphite crystals ($\sim\mathrm{20\,nm}$ thick). Metal leads were patterned by electron-beam lithography and subsequent deposition of metals (Cr 1~nm/ Au 50~nm). Monolayer $\mo$ flakes were transferred to this structure by dry viscoelastic stamping technique \cite{PDMS}. For more details see \SI.

\textbf{Optoelectronic Measurements}. To provide a source-drain bias the external DC source of a standard lock-in amplifier (SR830) was used. While to provide a gate bias the DC source of the lock-in amplifier or a Keithley 2400 were used. The current of the devices was collected by a pre-amplifier and then measured by a multimeter (Keithley 2000). To generate the photocurrent in the $\mo$ FET a 488~nm laser beam was focused in the devices by a 50$\times$ objective lens ($\sim\mathrm{1\,\mu m}$ spotsize).

\subsection*{Acknowledgments}
This work was supported by CAPES, Fapemig, CNPq, Rede de Nano-Instrumentação and INCT/Nanomateriais de Carbono.
The authors are thankful to the Laboratory of Nano Spectroscopy at UFMG for providing an experimental setup for this work, and to Centro Brasileiro de Pesquisas Fisicas (CBPF) and Centro de Componentes Semicondutores (CCS) for providing an e-beam lithography system and the Lab Nano at UFMG for allowing the use of an atomic force microscope.

\subsection*{Competing Financial Interests}
The authors declare no competing financial interests
\end{document}